\documentclass[10pt,aps,reprint,pra]{revtex4-1}   
\usepackage{amsmath}    
\usepackage{graphicx}   
\usepackage{bbold}
\usepackage{natbib}
\usepackage[usenames]{color}

\begin{document}

\title{Various quantum nonlocality tests with a simple 2-photon entanglement source}
\author{Enrico Pomarico}
\email{enrico.pomarico@unige.ch}
\author{Jean-Daniel Bancal}
\author{Bruno Sanguinetti}
\author{Anas Rochdi}
\author{Nicolas Gisin}
\affiliation{Group of Applied Physics, University of Geneva, 1211 Geneva 4, Switzerland}


\begin{abstract}
Nonlocality is a fascinating and counterintuitive aspect of Nature, revealed by the violation of a Bell inequality.
The standard and easiest configuration in which Bell inequalities can be measured has been proposed by Clauser-Horne-Shimony-Holt (CHSH).
However, alternative nonlocality tests can also be carried out.
In particular, Bell inequalities requiring multiple measurement settings can provide deeper fundamental insights about quantum nonlocality as well as offering advantages in the presence of noise and detection inefficiency. In this article we show how these nonlocality tests can be performed using a commercially available source of entangled photon pairs.
We report the violation of a series of these nonlocality tests ($I_{3322}$, $I_{4422}$ and chained inequalities). With the violation of the chained inequality with 4 settings per side we put an upper limit at 0.49 on the local content of the states prepared by the source (instead of 0.63 attainable with CHSH). We also quantify the amount of true randomness that has been created during our experiment (assuming fair sampling of the detected events).
\end{abstract}

\maketitle

\section{Introduction}
Nonlocality is one of the most counterintuitive and fascinating aspects of Nature revealed by quantum theory.
Indeed, the fact that two separated systems appear to work in a joint way, independently of the distance separating them, does not have a counterpart in the classical world. In particular, this bizarre effect is predicted by quantum mechanics for two entangled systems and manifest itself in the correlations of the outcomes of the measurements performed on the two systems.
This kind of "spooky actions at a distance" was the argument that Einstein, Podolski and Rosen used to claim the incompleteness of the quantum mechanics \cite{Einstein35}. They pointed out that one could restore locality in physics by assuming the existence of local variables determining the results of the measurements.
However, in 1964 John Bell showed that the correlations of the results of the measurement on entangled particles are stronger with respect to what is expected by any physical theory of local hidden variables. By measuring them, a simple inequality (i.e. the Bell inequality  \cite{Bell64}), based on the locality assumption, can be violated.

Testing experimentally nonlocal correlations predicted by quantum mechanics has become a concrete idea after the Clauser-Horne-Shimony-Holt (CHSH) formulation \cite{Clauser69} of the original Bell inequality.
Since the CHSH-Bell tests performed by Aspect in 1981 \cite{Aspect81}, several experiments have confirmed consistent violations of the CHSH inequality. However, the experimental imperfections, which all these tests are susceptible to, open loopholes that can be exploited by a local theory to reproduce the experimental data.
Two relevant loopholes are the detection and the locality loophole. The former relies on the fact that particles are not always detected in both channels of the experiment. The latter is related to the necessity of separating the two sites enough to prevent any light-speed communication between them from the measurement settings are set until the detection events have occurred.
At the moment both loopholes have been closed, but not yet within the same experiment.

Nowadays, performing nonlocality tests with entangled photons is relatively easy. Indeed, the technology necessary for generating entanglement is well known and entanglement sources can be extremely compact, cheap and stable in time. Moreover, polarization entanglement can be analyzed in a practical and accurate way. Therefore, the experimental study of entanglement is not confined to a few privileged laboratories, but can be also carried out as part of undergraduate laboratory courses.

The CHSH inequality requires a rather simple configuration: two binary measurement settings on each of two entangled particles.
However, in recent years, new types of Bell inequalities, involving a larger number of measurement settings or outcomes with respect to CHSH,
have been investigated from a theoretical point of view. These inequalities can get conclusions unattainable with CHSH: from a fundamental point of view, they allow for a deeper understanding of the nonlocal correlations, whereas from a practical one, they represent useful tools in the presence of noise and detection inefficiency \cite{Vertesi10}, as we show in the next paragraph.

In this paper, we show that, with a commercial source of photon pairs entangled in polarization (QuTools \cite{qutools}), a series of nonlocality tests alternative to CHSH can be made. In particular, we report the violation of Bell inequalities requiring multiple binary measurement settings. To our knowledge these inequalities (with the exception of $I_{3322}$ \cite{Altepeter05}) have not been measured before.

In section \ref{par:multiple_settings_imnequalities} we justify the interest in this kind of inequalities from a fundamental and practical point of view.
In section \ref{par:tomography_optimization} we describe the partial tomography of the density matrix of the states prepared by the source and the optimization of the settings for enhancing the violation of the Bell inequalities.
In section \ref{par:experimental_part} we report the violation of inequalities inequivalent to CHSH, in particular the $I_{3322}$ \cite{Collins04} and two $I_{4422}$ inequalities \cite{Brunner08}. Then, we show the violations of chained inequalities \cite{Braunstein90} from $3$ to $6$ settings per side.
In section \ref{par:intepretation_EPR2} we interpret the violation of the chained inequalities according to a specific nonlocality approach introduced by Elitzur, Popescu, and Rohrlich (EPR2) \cite{Elitzur92}. This allows us to put an upper limit on the local content of the prepared states that is stronger than the one attainable by CHSH. Finally, we show that the observed violations allow one to certify that true random numbers have been created during the experiment. Throughout this work we assume fair sampling of the detected events, which allows us to avoid detection loophole issues.

\section{Bell inequalities with multiple measurement settings}\label{par:multiple_settings_imnequalities}

\begin{figure}[t]
\begin{center}
\includegraphics[width=0.4\textwidth]{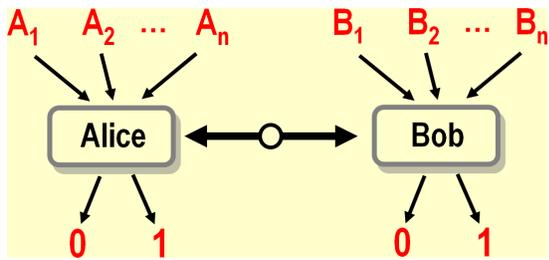}
\caption{Nonlocality tests where Alice and Bob measure the $N$ binary operators \{$A_1$,...,$A_N$\} and \{$B_1$,...,$B_N$\} respectively on the photon pairs produced by the entanglement source in the middle.}
\label{fig:inequality_scheme}
\end{center}
\end{figure}

A CHSH-Bell test requires the measurement of each photon of an entangled pair in two different bases and the estimation of the correlations in the four possible combinations of bases. Its practical implementation is conceptually easy and needs minimum experimental effort with respect to other nonlocality tests. Moreover, the CHSH test is particularly robust against the noise present in real experiments.
However, other Bell inequalities, requiring a larger number of measurement settings, schematically represented in Figure \ref{fig:inequality_scheme}, can lead to conclusions about the quantum correlations that are nontrivial and sometimes inaccessible to CHSH.
In general, finding all the Bell inequalities for a given setup is computationally hard, therefore the research is limited to a small number of settings per side.
Even in this case, these tests can provide interesting insights about quantum nonlocality.
In the case of three possible 2-outcome measurements per side, only one inequality, called $I_{3322}$, is inequivalent to CHSH \cite{Collins04}.
Note that this inequality is relevant in the sense that it can be violated by specific mixed 2-qubit states that do not violate CHSH. The $I_{3322}$ inequality has also been used to show that three qubits can share bipartite non-locality between more than two subsystems, a result that cannot be obtained with CHSH \cite{Collins04}.
In the case of four 2-outcome measurements per side, only a partial list of inequalities $I_{4422}$ has been given \cite{Brunner08}. Some of these inequalities are maximally violated, surprisingly, by non-maximally entangled states, unlike CHSH.

Another set of inequalities that have recently attracted attention is represented by the chained inequalities \cite{Braunstein90}, which are generalizations of CHSH with multiple settings.
In recent years, several theoretical models have attempted to provide a better understanding of quantum nonlocality.
One of these is the Elitzur, Popescu, and Rohrlich (EPR2) approach \cite{Elitzur92}, according to which the observed data could be explained assuming that only a fraction of the photon pairs produced in an experiment possesses nonlocal properties, while the remaining part gives rise to purely local correlations.
In this scenario chained inequalities are a useful tool for studying the local content of the quantum correlations. For instance, they have been used to prove that maximally entangled quantum states in arbitrary dimensions have a zero local component \cite{Barrett06} and to decrease the upper bound on the local content of nonmaximally entangled states \cite{Scarani06}.

From a more practical point of view, inequalities based on multiple settings are also interesting.
Recently, it has been shown that in the presence of high dimensional entanglement, that is when the quantum systems sharing entanglement have dimensions larger than two, these inequalities can tolerate a detection efficiency of 61.8\% for closing the detection loophole \cite{Vertesi10}. This value is lower with respect to the limit imposed by CHSH in experiments of entangled qubits.


\section{Optimization of the measurement settings for a specific state}\label{par:tomography_optimization}

Entangled states prepared in the laboratory (or by a commercial source) are not perfect in terms of purity and degree of entanglement. In this case the measurement settings needed to observe the largest possible violation of a given inequality do not necessarily coincide with those that are optimal for a maximally entangled state. In order to find these best measurement settings, a knowledge of the state is required.
Usually, a complete reconstruction of the state is not possible or even not necessary.
The QuTools source set-up \cite{qutools}, that we use in our experiment, projects the photons onto linear polarization states (c.f. description of the experimental setup in the next section), so we can perform a partial tomography of the state.
Once we know the state on the equatorial plane of the Bloch sphere corresponding to linear polarizations, we can optimize the measurement settings in this plane.

\subsection{Partial tomography of a quantum state} \label{par:partial_tomography}

The density matrix $\rho$ of a two-qubit state can always be written in the basis composed by the identity $\mathbb{1} $ and the Pauli matrices $\{\sigma_x, \sigma_y, \sigma_z\}$ as
\begin{eqnarray}\label{eq:rho}
\rho=\frac{1}{4}\Big(\mathbb{1}\otimes \mathbb{1} &+& \sum_{i=x,y,z} a_i \sigma_i \otimes \mathbb{1}+\sum_{j=x,y,z} b_j \mathbb{1} \otimes  \sigma_j\nonumber\\
& +& \sum_{i,j=x,y,z} c_{i,j} \sigma_i \otimes  \sigma_j \Big),
\end{eqnarray}
where $a_i=\langle \sigma_i \otimes \mathbb{1}\rangle_{\rho}$, $b_j=\langle \mathbb{1} \otimes  \sigma_j \rangle_{\rho}$ and $c_{i,j}=\langle \sigma_i \otimes  \sigma_j \rangle_{\rho} $ are $15$ real coefficients which completely define the state. A complete tomography \cite{James01} allows to determine the value of all these coefficients.

In our case we only measure linear polarizations, so a complete knowledge of the state is not necessary in order to predict all possible measurement statistics. Indeed, only the $8$ coefficients $a_i$, $b_j$, $c_{i,j}$ with $i,j=x,z$ are useful. These coefficients can be measured with our setup, realizing a partial tomography of the generated state.

\subsection{Optimization of the settings}\label{par:optimization_settings}

The measurement of a qubit along a particular angle $\theta$ in the $xz$ plane of the Bloch sphere can be represented by the measurement operator
\begin{equation} \label{eq:operator_plane}
O(\theta)=\cos\theta \sigma_z + \sin\theta \sigma_x.
\end{equation}
If the two-qubit state \eqref{eq:rho} is shared between Alice and Bob, the following marginal values are expected if Alice measures along angle $\alpha$ and Bob along $\beta$:
\begin{eqnarray}
E(\alpha) &= \langle A (\alpha) \otimes  \mathbb{1} \rangle_{\rho} = \cos\alpha a_z + \sin\alpha a_x,\\
E(\beta) &= \langle \mathbb{1} \otimes  B(\beta)\rangle_{\rho} = \cos\beta b_z+\sin\beta b_x.
\end{eqnarray}
Moreover, the joint correlations are found to be
\begin{eqnarray}
E(\alpha,\beta)&=&\langle A(\alpha) \otimes B(\beta) \rangle_{\rho}\nonumber\\
&=&\cos\alpha \cos\beta c_{z,z} + \cos\alpha \sin\beta c_{z,x}  \nonumber\\
&+&\sin\alpha \cos\beta c_{x,z} + \sin\alpha \sin\beta c_{x,x}.
\end{eqnarray}

A Bell inequality $I$ with $N$ settings per side \cite{Werner01} can be generally defined by the following formula
\begin{equation}\label{eq:definition_inequality}
I = \sum_{i=1}^{N} n_i E(\alpha_i)+\sum_{j=1}^{N} m_j E(\beta_j) + \sum_{i,j=1}^{N} l_{ij} E(\alpha_i,\beta_j) \leq I_L ,
\end{equation}
where $\alpha_i$ and $\beta_i$ are the angles of the measurements in Alice and Bob site respectively, $n_i$, $m_j$ and $l_{ij}$ are coefficients defining the inequality and $I_L$ is the local bound. The inequality $I$ can be represented schematically as a table
\begin{equation}
I=\left(
\begin{tabular}{c|cccc}
            & $m_1$    & $m_2$ &  ...  & $m_N$  \\
\hline
$n_1$         & $l_{11}$  & $l_{12}$&  ...  & $l_{1N}$ \\
$n_2$         & $l_{21}$  & $l_{22}$&  ...  & $l_{2N}$ \\
...           &  ...      &  ...    &  ...  &  ...     \\
$n_N$         & $l_{N1}$  & $l_{N2}$&  ...  & $l_{NN}$ \\
\end{tabular}
\right)\leq I_L.
\end{equation}
Note that this table has the same form as the one introduced in \cite{Collins04}, but here coefficients correspond to expectations values and not to probabilities.

It is clear from eq. (\ref{eq:definition_inequality}) that, for a given quantum state, the term $I$ is a function of the measurement angles on Alice's ($\{\alpha_1,...,\alpha_N\}$) and Bob's site ($\{\beta_1,...,\beta_N\}$), that is
\begin{equation}
I=I(\alpha_1,...,\alpha_N,\beta_1,...,\beta_N).
\end{equation}
This function is not linear in terms of its $2N$ variables, but numerical optimizations can be used in order to find the optimal measurement settings for Alice and Bob, which will provide the largest possible value of $I$ for the state under consideration.

For the nonlocality tests reported in this paper, we perform a partial tomography of the polarization entangled states that we prepare, as explained in the previous section. This allows us to reconstruct the state in the plane of the Bloch sphere corresponding to linear polarizations. Then, we numerically find the optimal measurement settings for this state, restricting them to lie on the $xz$ plane. These settings are used for the experimental violation of the Bell inequalities taken in consideration.\\

\section{Nonolocality tests with multiple settings}\label{par:experimental_part}

\subsection{The entanglement source}

\begin{figure}[h]
\begin{center}
\includegraphics[width=0.4\textwidth]{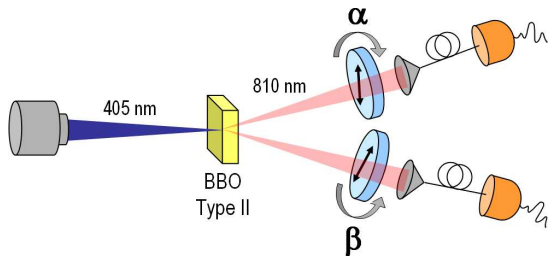}
\caption{Simple sketch of the polarization entanglement source. Linear polarizers at the Alice and Bob site are used for the settings necessary for the measurement of the Bell inequalities.}\label{fig:setup}
\end{center}
\end{figure}

We use the commercial entanglement source sold by the QuTools company \cite{qutools}, to which we brought some minor modifications. This source generates photon pairs entangled in polarization from Spontaneous Parametric Down Conversion (SPDC) in a bulk BBO crystal. The nonlinear crystal is cut for a Type II phase-matching and is pumped by a continuous wave diode laser at 405\,nm. Photon pairs at 810\,nm are generated, filtered and collected into single mode fibers.
Linear polarizers at the Alice and Bob site can measure polarizations respectively at the angles $\alpha$ and $\beta$ with respect to the vertical direction, as shown by the sketch in figure \ref{fig:setup}.
They allow to perform a partial tomography of the entangled state.
Notice that a full tomography could be done by just inserting quarter wave plates.
Photons are then detected using a Silicon Avalanche PhotoDiodes (Si-APDs) with efficiencies at the wavelength of the photons of 48\% and 55\% respectively. The coincidences are measured by a Time-to-Digital Converter (TDC) and a coincidence rate of 4.2\,kHz is measured for 15\,mW of input pump power in the absence of the linear polarizers.
Other details on this source can be found in \cite{Trojek04}.
For a pair of polarization directions measured by the two polarizers, the number of coincidences in a time interval of 20\,s is measured. The error associated to the coincidence rate is estimated according to poissonian statistics.
The inequalities that we want to test require to measure a large number of correlations. One correlation term is given by measuring one photon in one basis and the other photon in another one. Measuring one photon in one basis means projecting it into two orthogonal polarization directions. So the measurement of one correlation term require four different settings for the two polarizers.

\subsection{The characterization of the state}
We perform a partial tomography of the state generated by the source on the plane of the Bloch sphere corresponding to the linear polarizations. We determine the $8$ correlation values indicated in the section \ref{par:partial_tomography}, demanding to measure on each side the operators in polarization:
\begin{eqnarray*}
\sigma_z &\equiv&|H\rangle\langle H| - |V\rangle\langle V|,\\
\sigma_x&\equiv&|+\rangle\langle +| - |-\rangle\langle -|,\\
\mathbb{1}&\equiv&|H\rangle\langle H| + |V\rangle\langle V|\equiv|+\rangle\langle +| + |-\rangle\langle -|,
\end{eqnarray*}
where $|H\rangle$ and $|V\rangle$ correspond to the states of horizontal and vertical polarizations respectively and $|+(-)\rangle=\frac{1}{\sqrt{2}}(|H\rangle \pm |V\rangle)$. In the Table \ref{tab:expectation_values} a list of the measured expectation terms for the partial tomography of the prepared state is given.
\begin{table}[h]
\centering
\begin{tabular}{ccc}
Expectation term & Exp  & Th\\
\hline\hline
$\langle \sigma_z \otimes \sigma_z \rangle $ & $-0.9649 \pm 0.0012$  & -1\\
$\langle \sigma_x \otimes \sigma_x \rangle $ & $-0.9344 \pm 0.0017$ & -1 \\
$\langle \sigma_z \otimes \sigma_x \rangle $ & $0.1053 \pm 0.0045$ & 0 \\
$\langle \sigma_x \otimes \sigma_z \rangle $ & $-0.0201 \pm 0.0048$ & 0 \\
$\langle \sigma_z \otimes \mathbb{1} \rangle $ & $0.065 \pm 0.034$ & 0 \\
$\langle \sigma_x \otimes \mathbb{1} \rangle $ & $0.036 \pm 0.014$ & 0 \\
$\langle \mathbb{1} \otimes \sigma_z \rangle $ & $-0.078 \pm 0.020$ & 0 \\
$\langle \mathbb{1} \otimes \sigma_x \rangle $ & $-0.015 \pm 0.019$ & 0 \\
\end{tabular}
\caption{List of the expectation terms for the partial tomography of the prepared entangled state. The measured values (Exp) can be compared to the theoretical ones (Th) that are expected from a perfect singlet state.}\label{tab:expectation_values}
\end{table}

The error associated to a correlation term is obtained simply by statistical propagation of the error in the coincidences on which it depends. The marginal terms, the last four in the table, have been measured twice using two different bases on Bob's site. The difference between these two values reflects in errors which are larger with respect to the other expectation terms. Note that errors on the measured expectation terms in table \ref{tab:expectation_values} need to be considered in order to define from the partial tomography a density matrix which is definite positive. In some cases, in order to avoid the negativity of the density matrix attainable from a tomography, some techniques, such as the maximum likelihood estimation, need to be employed \cite{James01}.
It is evident that the singlet state prepared with the commercial source has some imperfections. There is a mixture component in the state since the expectation values are slightly different to the theoretical ones. The state is also unbalanced in the two orthogonal bases $\{|H\rangle, |V\rangle\}$ and $\{|+\rangle, |-\rangle\}$, which confirms that it is not maximally entangled.

The problem of optimization of the measurement settings for the nonlocality tests is necessarily limited to the plane on which we have limited the tomography.
We measure $9$ inequalities requiring in total 332 settings of the linear polarizers in the Alice and Bob site. These optimal settings have in some cases differences of some degrees with respect to the standard settings used for obtaining a maximum violation with a perfect singlet state.

\subsection{CHSH inequality}
First of all, in order to check the validity of our optimization method, we measure the CHSH inequality using the standard settings for the singlet state, then the same inequality with the settings optimized for the prepared state.
According to the notation given in \ref{par:optimization_settings}, we can represent the CHSH inequality by the following table
\begin{equation}
I_{CHSH}=\left(
\begin{tabular}{c|cc}
            & $0$    & $0$ \\
\hline
$0$         & $1$  & $1$\\
$0$         & $-1$  & $1$\\
\end{tabular}
\right)\leq 2.
\end{equation}
For the CHSH test with optimal settings we obtain a value of $2.731 \pm	0.015$, clearly enhanced with respect to $2.691\pm 0.015 $, obtained by adopting the standard settings. These two values correspond to violations of the local bound of respectively $49$ and $46$ standard deviations. This confirms the validity of the settings' optimization.
Actually, we expected from the partial tomography a value for CHSH of 2.683 with optimal settings and 2.662 with the standard ones. This little discrepancy could be explained by small variations of the prepared state between the tomography and the final measurements.

\subsection{Inequalities inequivalent to CHSH}
We then measure Bell inequalities inequivalent to CHSH, in particular $I_{3322}$ \cite{Collins04} and two different types of $I_{4422}$, called $AS_1$ and $AS_2$ \cite{Avis06,Brunner08}. These inequalities are facets of their corresponding Bell polytope \cite{Werner01}. They are thus optimal to detect nonlocality for some correlations in scenarios involving three and four settings. Note that $I_{3322}$ asks to measure not only joint correlations, but also four marginal probabilities. On the contrary, the two $I_{4422}$ inequalities that we want to measure are the only two of this kind to require correlation terms uniquely. This makes their measurement simpler from a conceptual and experimental point of view. In the following, the coefficients of these inequalities are given
\begin{equation}
I_{3322}=\left(
\begin{tabular}{c|ccc}
            & $1$  & $1$  & $0$\\
\hline
$1$         & $-1$  & $-1$  & $-1$\\
$1$         & $-1$  & $-1$  & $1$\\
$0$         & $-1$  & $1$ & $0$\\
\end{tabular}
\right)\leq 4,
\end{equation}
\begin{equation}
AS_1=\left(
\begin{tabular}{c|cccc}
            & $0$  & $0$  & $0$ & $0$\\
\hline
$0$         & $1$  & $1$  & $1$ & $1$\\
$0$         & $1$  & $1$  & $1$ & $-1$\\
$0$         & $1$  & $1$ & $-2$ & $0$\\
$0$         & $1$  & $-1$ & $0$ & $0$\\
\end{tabular}
\right)\leq 6,
\end{equation}
\begin{equation}
AS_2=\left(
\begin{tabular}{c|cccc}
            & $0$  & $0$  & $0$ & $0$\\
\hline
$0$         & $2$  & $1$  & $1$ & $2$\\
$0$         & $1$  & $1$  & $2$ & $-2$\\
$0$         & $1$  & $2$ & $-2$ & $-1$\\
$0$         & $2$  & $-2$ & $-1$ & $-1$\\
\end{tabular}
\right)\leq 10.
\end{equation}
Note that this notation is different with respect to that used in \cite{Brunner08}.
We call $I^{exp}$ the experimental value of the Bell parameter. For each of the three different inequalities, we observe a violation of the local bound $I_L$ (Table \ref{tab:inequalities_values}). In the Table \ref{tab:inequalities_values} the result for CHSH is that with optimal settings. The violations for the two $I_{4422}$ inequalities are stronger than for $I_{3322}$. We evaluate this aspect by considering the resistance to noise of the three different inequalities.
If some white noise were added with probability $p_{noise}$ to our state $\rho$, we would obtain a state with a density matrix $\rho_{noise}=p_{noise}\frac{\mathbb{1}}{4}+(1-p_{noise}) \rho$. Now, it is easy to show that the violation would not be observed anymore if $p_{noise}>1-\frac{I_L}{I^{exp}}$. For $I_{3322}$ adding $13\%$ of white noise to the state compromises the violation. For the other inequalities the tolerance to the noise is higher and the CHSH inequality confirms to be the most robust to noise.

In all the cases the obtained violations agree quite well with what we expected with the tomography.

\begin{table}[h]
\centering
\begin{tabular}{c|ccccc}
& $I_{L}$ & $I^{exp}$ &  $I^{tom}$ & $I^{exp} - I_{L}$ & $p_{\text{noise}} (\%)$ \\
&          &          &            &($\sigma$ units)   &  \\
\hline\hline
$I_{CHSH}$ & 2  & 2.731	$\pm$	0.015   & 2.683 & 49 & 27 \\
$I_{3322}$ & 4  & 4.592	$\pm$	0.024   & 4.769 & 25 & 13  \\
$AS_1$   & 6    & 7.747	$\pm$	0.026   & 7.750 & 67 & 23\\
$AS_2$   & 10   & 12.85	$\pm$	0.030   & 12.819& 95 & 22 \\

\hline
\end{tabular}
\caption{Measurement of the CHSH inequality and of the inequalities inequivalent to CHSH. $I_L$ is the local bound, $I^{exp}$ is the value of the Bell parameter obtained experimentally with the optimized settings, $I^{tom}$ is the expected value from the partial tomography, $I^{exp} - I_{L}$ is the difference between the obtained value and the local bound in terms of number of standard deviations $\sigma$ and $p_{\text{noise}} (\%)$ is the critical level of white noise that can be added to the system while still keeping a violation. }\label{tab:inequalities_values}
\end{table}

\subsection{Chained inequalities}
Finally, we measure chained inequalities with a number of settings per side from $2$ to $6$.
The chained inequality with $N$ settings for Alice ($\{\alpha_1,\alpha_2,...,\alpha_N\}$) and $N$ for Bob ($\{\beta_1,\beta_2,...,\beta_N\}$) can be written as a correlation inequality as
\begin{eqnarray}
I^{(N)}_{chain}&=&E(\alpha_1,\beta_1)+E(\beta_1,\alpha_2)+E(\alpha_2,\beta_2)+...+\nonumber\\
&&+E(\alpha_N,\beta_N)-E(\beta_N,\alpha_1)\leq 2(N-1).
\end{eqnarray}
Contrary to the previous inequalities, only the chained inequality with $N=2$, which is equivalent to CHSH, is a facet of the Bell polytope.

Since the chained inequalities require $2N$ correlation terms, the number of settings required for their measurement scales linearly with N. We thus limit ourselves to $6$ settings per side. In the Table \ref{tab:chained_inequalities} a list of the measured inequalities is given. For $N=2$ we have the result for the CHSH inequality with optimal settings. All the inequalities are violated in a way consistent with the expected results and with a large number of standard deviations of difference with respect to the local bound. However, it is interesting to note that the larger the number of settings, the weaker the violation. Indeed, the fraction of noise that we can add to the system and still keep the violations decreases for an increasing number of settings.

\begin{table}[h]
\centering
\begin{tabular}{c|ccccc}
N   & $I_{L}$ & $I^{exp}$ &  $I^{tom}$& $I^{exp} - I_L$ & $p_{noise}$ (\%) \\
    &         &           &  &($\sigma$ units)  &  \\
\hline\hline

2   & 2  & 2.731	$\pm$	0.015   & 2.683 & 49 & 27 \\

3   & 4  & 4.907	$\pm$	0.019  &  4.925 & 48 & 18   \\

4   & 6  & 7.018	$\pm$	0.023   & 6.999 & 44 & 15  \\

5   & 8  & 8.969	$\pm$	0.026   & 8.996 & 37 & 11\\

6   & 10 & 10.91	$\pm$	0.028  & 10.954 & 33 & 8 \\

\hline
\end{tabular}
\caption{Measurement of the chained inequalities with $N$ settings per side.}\label{tab:chained_inequalities}
\end{table}

\section{Application of the chained inequalities}

As recalled in the section \ref{par:multiple_settings_imnequalities}, the violation of the chained inequalities can be linked with numerous problems. In the following we briefly mention what the violations we observed allow us to conclude with respect to the EPR2 model of nonlocality and to true randomness.

\subsection{EPR2 nonlocality}\label{par:intepretation_EPR2}

Chained inequalities can be used to put an upper bound on the local content of the prepared state according to the EPR2 approach \cite{Elitzur92}.
Indeed, as already explained in the section \ref{par:multiple_settings_imnequalities}, one can imagine that only part of the photon pairs produced by the source has nonlocal properties, while the other one behaves in a purely local way.
In such a situation, the measured value of $I_{chain}^{(N)}$ decomposes as a sum of the local bound $I_L$ with probability $p_L$ and of some possibly larger value $I$ with probability $1-p_L$. The latter satisfies the no-signaling (NS) principle, i.e. the impossibility of faster-than-light communication. Therefore
\begin{equation}\label{eq:EPR2}
I^{exp} = p_{L} I^{(N)}_L + (1-p_L) I_{NS}^{(N)}.
\end{equation}

Since the local and the no-signaling values of the $I$ quantity are at most
$I^{(N)}_L=2(N-2)$ and $I_{NS}^{(N)}=2N$ respectively, the local part of the produced
state is bounded by the maximum value $p_L^{max}$ :
\begin{equation}\label{eq:pl}
p_L\leq p_L^{max} = N-\frac{I^{exp}}{2}.
\end{equation}

To illustrate this bound, let us see how it applies to noisy singlet states, i.e. Werner states $\rho_{W}=V|\psi^{-}\rangle\langle \psi^{-}| + (1-V)\frac{\mathbb{1}}{4}$. For these states the maximum value of $I^{(N)}$ is $2 N V\cos(\frac{\pi}{2N})$.
\begin{figure}[h]
\begin{center}
\includegraphics[width=0.47\textwidth]{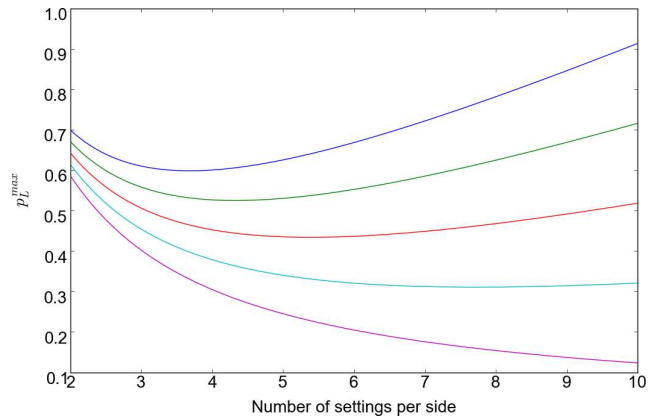}
\caption{Largest local part associated to a Werner state for different values of $V=1,0.98,0.96,0.94,0.92$ (from the bottom to the top).}\label{fig:pL_Werner}
\end{center}
\end{figure}
Thus, the maximum local probability $p_L^{max}$ associated to a Werner state $\rho_{W}$ depends on the visibility $V$ of the state and it is given by $p_L^{max}=N(1-V \cos(\frac{\pi}{2N}))$. The behaviour of this function is shown in the figure \ref{fig:pL_Werner} for different values of the visibility $V$.
For the pure singlet state ($V=1$), $p_L^{max}$ tends to $0$ for an infinite number of settings, confirming the fact that the singlet state is fully nonlocal \cite{Barrett06}. On the contrary, when $V$ is smaller than $1$, $p_L^{max}$ decreases until a certain limit value. Therefore, the number of settings which are useful for lowering the upper bound on the local content of the state is limited and changes according to the visibility $V$.

\begin{figure}[h]
\begin{center}
\includegraphics[width=0.47\textwidth]{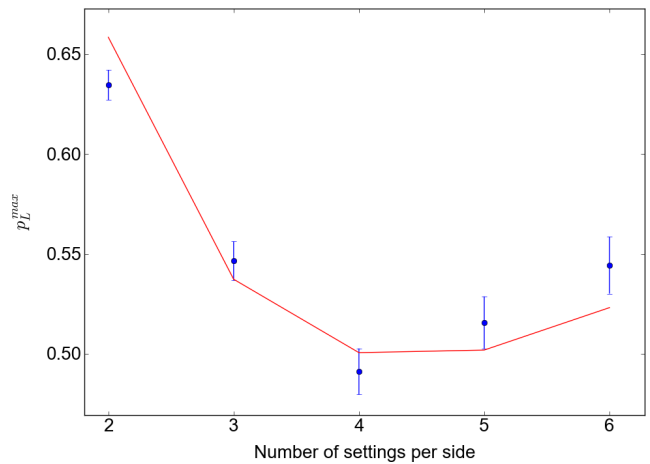}
\caption{Largest local part as a function of the number of settings per side. The red line joins the expected values for our state. The minimum measured value of $p_L^{max}$ is $0.491 \pm 0.012$ with a chained inequality of $4$ settings per side.}\label{fig:plocal}
\end{center}
\end{figure}

We observe a similar effect for the state produced by our source. In figure \ref{fig:plocal} we represent the violations of the measured chained inequalities in terms of local probabilities, as calculated from \eqref{eq:pl}. The red line joins the expected values for $p_L^{max}$ obtained by the knowledge of the state. Our state is more complex than a Werner state but it can be approximated by a Werner state with visibility $V=0.94$. The minimum value of $p_L^{max}$ that we measure is $0.491 \pm 0.012$ with a chained inequality of $4$ settings per side. This result means that at least half of the photon pairs produced by the source are nonlocal, a result that cannot be shown with the CHSH inequality. To our knowledge, this is the first experiment that fixes an upper bound on the local part of the quantum correlations according to the EPR2 approach \footnote{Note that existing experimental results could be reinterpreted to provide such a bound as well.}. Similar work in progress is expected to show a lower value of this quantity \cite{Aolita11}.

\subsection{Randomness certified by the no-signaling principle}

When measuring a singlet state, locally random outcomes can be observed. It has been recently shown that violation of a Bell inequality can certify that this randomness truly emerges during the experiment, in the sense that no algorithm can possibly predict the measured outcomes \cite{Pironio10}. Indeed, if such algorithm existed prior to the experiment, it could be considered as a local hidden variable, and no violation of a Bell inequality can be observed with only local hidden variables.
Note that in order to discard any such algorithm, the detection loophole should be closed during the experiment. Indeed, such an algorithm
could in principle inform detectors when they should click according to some detection loophole model \cite{Gisin02}.
In order to avoid this issue, we assume fair sampling of the detected events: the state of the particles coming onto a detector does not affect the fact that a detector fires or not, so that the detected pairs of particles fairly represent the ones produced by the source.

In order to quantify the amount of true randomness that can be found in some experimental results, one must consider all the marginal probabilities $P(a|x)$ that Alice finds the outcome $a$ when she measures $x$, which are compatible with the observed Bell inequality violation $I^{exp}$. We searched for the largest marginal probability $P(a|x)$ numerically among all possible quantum correlations, i.e. correlations that can be achieved by measuring a quantum state, as well as among all no-signaling correlations, denoting this largest quantity by $P^*(A|X)$.

\begin{figure}[h]
\begin{center}
\includegraphics[width=0.53\textwidth]{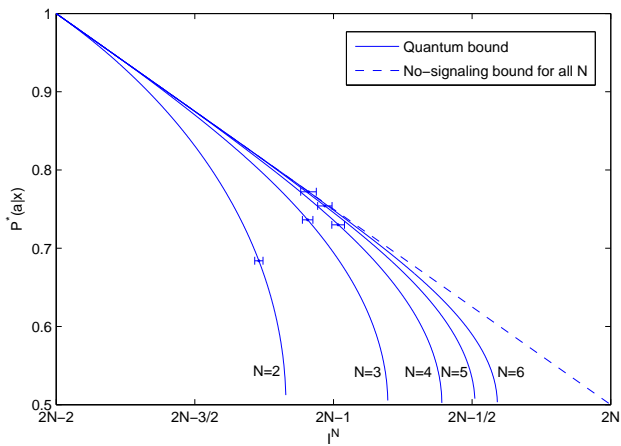}
\caption{Upper-bound on the marginal probability distribution as a function of the inequalities violation. The bound implied by the no-signaling principle is identical for all chained inequalities.}\label{fig:localRandomness}
\end{center}
\end{figure}

The computed upper bounds $P^*(A|X)$ for the chained inequalities with up to 6 settings per party are shown in figure \ref{fig:localRandomness} together with our experimental results.
The lowest bound on the marginal probability that we can certify here is $P^*(A|X)=0.684\pm0.014$, achieved for the CHSH inequality. However, if we consider the bound imposed by the no-signaling principle only, then the strongest one is $P^*(A|X)=0.7455\pm0.0057$, achieved by the chained inequality with $N=4$ settings per party.

Note that in order to extract a truly random bit string out of measured outcomes, classical key distillation techniques should be used \cite{Ronen02}. The ratio between the number of measured bits and the number of truly random, uniformly distributed, bits that is produced by this procedure is given by the min-entropy of Alice's outcome $A$ conditioned on her measurement choice $X$: $H_{\text{min}}(A|X)=-\log_2 \max_a P^*(A|X)$. In our case we find $H_{\text{min}}=0.55\pm0.03$ for the CHSH violation, meaning that approximately one random bit every two measurements have been created.

\section{Conclusions}
We have shown that non trivial nonlocality tests alternative to CHSH can be performed even with a rather simple commercial source. In particular, we have measured Bell inequalities with multiple measurement settings.
In particular, we have reported violations of $I_{3322}$, of two $I_{4422}$ and of chained inequalities with up to $6$ settings per side. Violations of these inequalities have not been shown before. Moreover, by using the chained inequalities, we have put an upper bound on the local content of the prepared state at $0.491 \pm 0.012$, meaning that at least half of the photon pairs produced by the source have nonlocal correlations. We have also quantified the amount of true randomness created in the experiment.
Therefore, even with an extremely simple setup, it is possible to implement not trivial nonlocality tests and get interesting conclusions on the nonlocal properties of the source and on the randomness produced in the experiment.
This work emphasizes the richness of nonlocality and the importance of the no-signaling and could be of interest for undergraduate laboratory courses.

\section{Acknowledgments}
This work is supported by EU project Q-ESSENCE and by the Swiss NCCR-Quantum Photonics.
We would like to thank Antonio Acin, Valerio Scarani and Henning Weier for valuable discussions, suggestions and remarks.


\begin{thebibliography}{20}
\bibitem{Einstein35}
A.~Einstein, B.~Podolsky, and N.~Rosen, "Can Quantum-Mechanical Description of Physical Reality Be Considered
  Complete?", Phys. Rev. \textbf{47}, 777 (1935).

\bibitem{Bell64}
J.~S.~Bell, "On the Einstein-Podolski-Rosen paradox", Phys. \textbf{1}, 195 (1964).

\bibitem{Clauser69}
J.~F.~Clauser, M.~A.~Horne, A.~Shimony, and R.~A.~Holt, "Proposed Experiment to Test Local Hidden-Variable Theories", Phys. Rev. Lett. \textbf{23}, 880 (1969).

\bibitem{Aspect81}
A.~Aspect, P.~Grangier, and G.~Roger, "Experimental Tests of Realistic Local Theories via Bell's Theorem",
Phys. Rev. Lett. \textbf{47}, 460 (1981).

\bibitem{Vertesi10}
T.~V\'{e}rtesi, S.~Pironio, and N.~Brunner,
"Closing the Detection Loophole in Bell Experiments Using Qudits",
Phys. Rev. Lett. \textbf{104}, 060401 (2010).

\bibitem{qutools}
http://www.qutools.com.

\bibitem{Altepeter05}
J.~B.~Altepeter, E.~R.~Jeffrey, P.~G.~Kwiat, S.~Tanzilli, N.~Gisin, and A.~Ac\'in,
"Experimental Methods for Detecting Entanglement",
Phys. Rev. Lett. \textbf{95}, 033601 (2005).

\bibitem{Collins04}
D.~Collins, and N.~Gisin,
"A relevant two qubit Bell inequality inequivalent to the CHSH inequality",
Journ. of Phys. A: Math. and Gen. \textbf{37}, 1775 (2004).

\bibitem{Brunner08}
N.~Brunner, and N.~Gisin, "Partial list of bipartite Bell inequalities with four binary
  settings",
Phys. Lett. A \textbf{372}, 3162 (2008).

\bibitem{Braunstein90}
S.~Braunstein, and C.~Caves,
"Wringing out better Bell inequalities",
Annals of Phys. \textbf{202}, 22 (1990).

\bibitem{Elitzur92}
A.~Elitzur, S.~Popescu, and D.~Rohrlich,
"Quantum nonlocality for each pair in an ensemble",
Phys. Lett. A \textbf{162}, 25 (1992).

\bibitem{Barrett06}
J.~Barrett, A.~Kent, and S.~Pironio,
"Maximally Nonlocal and Monogamous Quantum Correlations",
Phys. Rev. Lett.  \textbf{97}, 17 (2006).

\bibitem{Scarani06}
V.~Scarani,
"Local and nonlocal content of bipartite qubit and qutrit
  correlations",
Phys. Rev. A \textbf{77}, 042112 (2008).

\bibitem{James01}
D.~F.~V.~James, P.~G.~Kwiat, W.~J.~Munro, and A.~G.~ White,
"Measurement of qubits",
Phys. Rev. A \textbf{64}, 052312 (2001).

\bibitem{Werner01}
R.~F.~Werner, and M.~M.~Wolf,
"Bell inequalities and Entanglement", Quant. Inf. Comp. \textbf{1}, 1 (2001).

\bibitem{Note1}
Note that existing experimental results could be reinterpreted to provide such a bound as well.

\bibitem{Aolita11}
L.~Aolita \textit{et al.}, in preparation.

\bibitem{Trojek04}
P.~Trojek, C.~Schmid, M.~Bourennane, H.~Weinfurter, and C.~Kurtsiefer,
"Compact source of polarization-entangled photon pairs",
Opt. Expr. \textbf{12}, 276 (2004).

\bibitem{Avis06}
D. Avis, H. Imai, and T. Ito,
"On the relationship between convex bodies related to correlation experiments with dichotomic observables",
Journ. of Phys. A: Math. and Gen. \textbf{39}, 11283 (2006).

\bibitem{Pironio10}
S.~Pironio, A.~Ac\'in, S.~Massar, A.~Boyer de~la Giroday, D.~N.
  Matsukevich, P.~Maunz, S.~Olmschenk, D.~Hayes, L.~Luo, T.~A. Manning, and
  C.~Monroe, "Random numbers certified by Bell's theorem", Nat. \textbf{464}, 1021 (2010).

\bibitem{Gisin02}
N.~Gisin, and B.~Gisin, "A local variable model for entanglement swapping
exploiting the detection loophole",
Phys. Lett. A \textbf{297} 279–284,(2002).

\bibitem{Ronen02}
S.~Ronen, "Recent developments in extractors",
Bulletin of the European Association for Theoretical Computer Science \textbf{77}, 67 (2002).

\end{thebibliography}

\end{document}